\documentclass[usenatbib]{mn2e}
\usepackage[dvips]{graphicx}
\usepackage{amssymb,amsmath}
\title[Luminous satellite galaxies in gravitational lenses]{Luminous satellite galaxies in gravitational lenses}
\author[S. E. Bryan et al.]
{S. E. Bryan$^{}$\thanks{E-mail:sarah.bryan@manchester.ac.uk}, 
S. Mao$^{}$ and
S. T. Kay$^{}$\\
Jodrell Bank Centre for Astrophysics, Alan Turing Building, University of Manchester, Manchester M13 9PL, U.K.\\
}
\begin{document}

\def\aj{AJ}					
\def\araa{ARA\&A}				
\def\apj{ApJ}					
\def\apjl{ApJL}					
\def\apjs{ApJS}					
\def\apss{Astrophysics and Space Science}
\def\capsp{Comments on Astrophysics and Space Physics}
\def\aap{A\&A}					
\def\aapr{A\&A~Rev.}				
\def\aaps{A\&AS}				
\def\azh{AZh}					
\def\baas{BAAS}					
\def\jrasc{JRASC}				
\def\memras{MmRAS}				
\def\mnras{MNRAS}					
\def\pasp{PASP}					
\def\pasj{PASJ}					
\def\qjras{QJRAS}				
\def\skytel{S\&T}				
\def\solphys{Sol.~Phys.}			
\def\sovast{Soviet~Ast.}			
\def\ssr{Space~Sci.~Rev.}			
\def\zap{ZAp}					
\def\na{New Astronomy}				
\def\iaucirc{IAU~Circ.}				
\def\aplett{Astrophys.~Lett.}			
\def\apspr{Astrophys.~Space~Phys.~Res.}		
\def\bain{Bull.~Astron.~Inst.~Netherlands}	
\def\memsai{Mem.~Soc.~Astron.~Italiana}		

\def\ao{Appl.~Opt.}				

\def\pra{Phys.~Rev.~A}				
\def\prb{Phys.~Rev.~B}				
\def\prc{Phys.~Rev.~C}				
\def\prd{Phys.~Rev.~D}				
\def\pre{Phys.~Rev.~E}				
\def\prl{Phys.~Rev.~Lett.}			
\def\nat{Nature}				
\def\fcp{Fund.~Cosmic~Phys.}			
\def\gca{Geochim.~Cosmochim.~Acta}		
\def\grl{Geophys.~Res.~Lett.}			
\def\jcp{J.~Chem.~Phys.}			
\def\jgr{J.~Geophys.~Res.}			
\def\jqsrt{J.~Quant.~Spec.~Radiat.~Transf.}	
\def\nphysa{Nucl.~Phys.~A}			
\def\physrep{Phys.~Rep.}			
\def\physscr{Phys.~Scr}				
\def\planss{Planet.~Space~Sci.}			
\def\procspie{Proc.~SPIE}			
\def\rpp{Rep.~Prog.~Phys.}			
\let\astap=\aap
\let\apjlett=\apjl
\let\apjsupp=\apjs
\let\applopt=\ao
\let\prep=\physrep


\date{Accepted ...... Received ...... ; in original form......   }

\pagerange{000 -- 000} \pubyear{0000}

\maketitle
\label{firstpage}
\begin{abstract}
Substructures, expected in cold dark matter haloes, have been proposed to explain the anomalous flux ratios in gravitational lenses.
About 25\% of lenses in the Cosmic Lens All-Sky Survey (CLASS) appear to have
luminous satellites within \mbox{$\sim$ 5 $ h^{-1}$ kpc} of the main lensing galaxies, which are usually at redshift $z\sim 0.2-1$.  In this work we use the Millennium Simulation combined with galaxy catalogues from semi-analytical techniques to study the predicted frequency of such satellites in simulated haloes.  The fraction of haloes that host bright satellites within the (projected) central regions is similar for red and blue hosts and is found to increase as a function of host halo mass and redshift.
Specifically, at $z = 1$, about $11$\% of galaxy-sized haloes (with masses between $10^{12} h^{-1}$ M$_\odot $ and $ 10^{13} h^{-1}$ M$_\odot$) host bright
satellite galaxies within a projected radius of 5 $ h^{-1}$ kpc.  This fraction increases to about 17\% (25\%) if we consider bright (all) satellites of only group-sized haloes (with masses between $10^{13} h^{-1}$ M$_\odot $ and $10^{14} h^{-1}$ M$_\odot$).  These results are roughly consistent with the fraction ($\sim 25\%$) of CLASS lensing galaxies observed to host luminous satellites.  At \mbox{$z = 0$}, only $\sim 3$\%
of galaxy-sized haloes host bright satellite galaxies.  The fraction rises to $\sim 6\%\, (10\%)$ if we consider bright (all) satellites of only group-sized haloes at $z = 0$.  However, most of the satellites found in the inner regions are `orphan' galaxies where the dark matter haloes have been completely stripped.  Thus the agreement crucially depends on the true survival rate of these `orphan' galaxies.  We also discuss the effects of numerical resolution and cosmologies on our results.
\end{abstract}

\begin{keywords}
cosmology: galaxy formation - gravitational lensing
\end{keywords}

\section{Introduction}

In the hierarchical scenario of structure formation, structures in the Universe are assumed to have grown from tiny quantum fluctuations (generated during an inflationary period) through gravitational instability.  Due to the shape of the power spectrum, structures form hierarchically - larger structures form via accretion and merging of smaller structures.  Dense cores of the smaller structures often survive the merging process and manifest as subhaloes in the primary haloes.  If substantial star formation occurs in these subhaloes (or their progenitors), then they will appear as satellite galaxies. 

In the Milky Way, hundreds of subhaloes are predicted, starting from
earlier semi-analytical studies (\citealt{bib:Kauffmann93}), to more
recent high-resolution simulations (\citealt{bib:Klypin99};
\citealt{bib:Moore99}; \citealt{bib:Gao04_gal,bib:Gao04_dm};
\citealt{bib:Diemand07}).  A few years ago, there were only a dozen or
so satellites known, far fewer than the predicted number of subhaloes.  However, very
recently, a new population of satellites has been discovered in the
Sloan Digital Sky Survey data (e.g. \citealt{bib:Belokurov07}).  It should be noted though that these satellites are compact and, in general, much fainter than the previously known ones, thus it is likely that even this new population of satellite galaxies cannot completely remove the discrepancy between simulations and observations (\citealt{bib:Madau08}).  It is possible that many subhaloes are dark due to inefficient star formation, for example, due to its suppression by the UV-background radiation (e.g. \citealt{bib:Doroshkevich67}; \citealt{bib:Couchman86}; \citealt{bib:Efstathiou92}). 
 
Such dark substructure can, potentially, be detected through several means, for example through gamma-ray radiation due to annihilations of dark matter particles (\citealt{bib:Stoehr03}; \citealt{bib:Die07}).  Gravitational lensing is, in principle, another way to detect dark (and luminous) substructure.  Flux anomalies (\citealt{bib:Mao98}), astrometric perturbations (\citealt{bib:Chen07}) and time-delays (\citealt{bib:Keeton08}) can be used to infer the presence of substructure in strong gravitational lenses.  The results of these studies are so far inconclusive (e.g. \citealt{bib:Kochanek04, 
2004ApJ...604L...5M}). If all substructure is equally efficient in affecting the flux ratios, then it is clear that there is more than sufficient mass in subhaloes to explain the flux anomalies.  Unfortunately, most subhaloes are in the outer part of the galaxy halo, which means they will have relatively little impact on the flux anomalies occurring in the central parts of lensing galaxies.  Curiously, as emphasised by Schneider (2007, private communication), 3 of the 6 radio lenses studied by \cite{bib:Kochanek04} exhibit luminous satellite galaxies close to the primary lensing galaxy, namely MG0414+0534, B1608+656 and B2045+265.  A question naturally arises: are such luminous satellite galaxies expected this frequently in the current structure formation theory?

The lensing cross-section is
dominated by elliptical galaxies, thus for lensing applications it is
important to divide galaxies into different types, and see whether the
subhalo populations are different.
Furthermore, we explore in more detail the evolution of satellite
galaxies as a function of redshift.  If the evolution is slow, we
can more conveniently use studies of nearby galaxies to infer the properties of the luminous
satellite population of galaxies at intermediate redshift (between 0.5 and 1, where
most lensing galaxies lie).  These are the two specific aspects of the
subhalo population that we will address in this paper.  For this purpose, we will use the largest cosmological simulation combined with semi-analytical catalogues to select haloes and study their satellite populations.  We compare our results to the CLASS survey (\citealt{bib:Browne03}; \citealt{bib:Myers03}). 

The plan of the paper is as follows.  In Section \ref{sims}, we describe the Millennium Simulation and the semi-analytical galaxy catalogue we use.  Our main results are presented in Section \ref{results}, and we finish with a summary in Section \ref{summary}.
\section{Numerical Simulation Data}
\label{sims}

\subsection{Millennium Simulation}

The Millennium Simulation, run by the Virgo Consortium\footnote{www.virgo.dur.ac.uk}, follows the evolution of $2160^{3}$ particles within a comoving box of length 500 $h^{-1}$ Mpc with a force softening length of 5 $h^{-1}$ kpc (where the Hubble constant $H_0 = 100 h$ km s$^{-1}$ Mpc$^{-1}$).  A $\Lambda$CDM cosmology is assumed, using parameters consistent with the results obtained from the first-year \emph{Wilkinson Microwave Anisotropy Probe} ({\emph{WMAP}}) data  \citep{bib:wmap1_03}: $\Omega_m$ = 0.25, $\Omega_\Lambda$ = 0.75, $h$ = 0.73, \mbox{$n$ = 1} and $\sigma_8$ = 0.9.  
Haloes are identified using the SUBFIND algorithm to detect self bound groups of at least 20 particles.  This implies a minimum mass for a detected halo of $1.7 \times 10^{10} h^{-1}$ M$_\odot$.  For a full description of the simulation the reader is referred to \cite{bib:Springel05}\footnote{http://www.mpa-garching.mpg.de/galform/virgo/millennium}. 
 \ \\
\subsection{Semi-analytic galaxy catalogues}

There are three publicly available galaxy catalogues that have been created using the Millennium Simulation to trace the underlying dark matter (\citealt{bib:Bower06,bib:Bertone07,bib:DeLucia07}).  We have used the \cite{bib:DeLucia07} galaxy catalogue for this analysis.  It is based on the model described in \cite{bib:Springel05} and \cite{bib:Croton06} and is similar to the techniques employed by \cite{bib:Springel01} and \cite{bib:DeLucia04}.  The catalogues created by \cite{bib:DeLucia07} and \cite{bib:Bertone07} are based on similar semi-analytic models, differing only in their implementation of galactic feedback.  We expect that the results we present here would not be significantly affected by the differences in the models.  The third model, produced by \cite{bib:Bower06}, does not differentiate between `orphan' galaxies and galaxies that are still associated with their dark matter haloes.\\
\ \\
For each halo identified in the simulation, a central galaxy is `created' with a mass fraction in baryons of 17\% (corresponding to the global ratio, $\Omega_b / \Omega_m$, as measured by {\emph{WMAP}} first-year data).  Initially the `created' galaxy has no stellar mass, no cold gas and zero luminosity.  The attributed baryons are in the form of diffuse gas with primordial composition.  The semi-analytic model uses merger trees taken from the Millennium Simulation to describe the evolution of haloes that host the galaxies.  The formation and evolution of the galaxies is then followed by implementing simple physical prescriptions for the baryonic physics, such as gas cooling, star formation and feedback processes (including AGN feedback).  These processes depend on the properties of the host halo.  When two haloes merge, the galaxy associated with the larger halo remains the central galaxy while the galaxy attributed to the lower mass progenitor becomes a satellite.  Satellite galaxies are stripped of their hot gas and have no new supply of cool gas.  They are allowed to form stars until their cool gas reservoir is exhausted.  Subhaloes are followed, after merging with a larger system, until the dark matter subhalo is completely disrupted by tidal forces.  These tidally stripped `orphan' galaxies are then assumed to follow the position of the most bound particle in the subhalo before it was disrupted, until it merges with the central galaxy on the dynamical friction timescale.  More details on the formation and evolution of the galaxies can be found in \cite{bib:Croton06}. \\
\ \\
 We note that, as the minimum mass of a resolved halo is  $1.7 \times 10^{10} h^{-1}$ M$_\odot$ and the creation of a galaxy relies on the existence of a halo, the number of low mass galaxies may well be underestimated.   This is unlikely to affect our results as the low mass satellites will have lower circular velocities and are likely to be faint.  This is discussed in more detail in Section \ref{discussion}.

\subsection{Galaxy sample}
\label{sec:sample}

We study the satellite population within massive galaxy-sized haloes (haloes with a virial mass\footnote{The mass enclosed within the virial radius. (We define the virial radius to be the radius of a sphere, centred on the most bound particle, within which the average density is 200 times the critical density).} between $10^{12} \, h^{-1}  \mbox{M}_\odot $  and $10^{13} \, h^{-1} $ M$_\odot$) and group-sized haloes (with $10^{13} \, h^{-1}  \mbox{M}_\odot \leq $ \mbox{M$_{vir}$ $ < 10^{14} \, h^{-1} $ M$_\odot$}).  We do not consider cluster-sized haloes as none of the CLASS lenses are found in such environments.  The number of galaxy-sized host haloes considered in this analysis is around  $3 \times 10^5$, varying little between $z = 0$ and $z = 1$.  For group-sized haloes, the number is around $3 \times 10^4$ at $z = 0$ decreasing to $\sim 2 \times 10^4$ at $z = 1$.  As in \cite{bib:Sales07}, we have imposed a brightness cutoff of \mbox{M$_R < -20.5$} on our central (host) galaxies to ensure that they have a reasonable chance of hosting detectable satellites. In any case, faint central galaxies have small lensing cross-sections, and will have little effect on the statistics (see Section \ref{sec:predictions}). We consider all galaxies (within the virial radius of their host) with M$_R < -17$  as luminous satellites. This corresponds, approximately, to a 100 particle halo - the morphological resolution limit of the simulation (see \citealt{bib:Croton06}).  This cut would not exclude any of the observed luminous satellites within the CLASS sample.
Furthermore, to aid in direct comparison with observation, bright satellites are required to have R-band luminosities between 1\% and 50\% of that of their host.  For completeness, we also consider fainter satellites by dropping the $R$-band magnitude and lower luminosity ratio cuts on our satellite sample.

Since the lensing cross-section is dominated by massive, red elliptical galaxies, we wish to divide our sample by galaxy type and determine
whether there is a significant difference in the subhalo population of red and blue galaxies.  To do this, we divide our according to the $B-V$ colour of the host galaxy.  The galaxy populations are bimodal as a function of colour, with a well-defined red sequence and a blue cloud (see e.g. \citealt{bib:Croton06};  their Fig. 9).  We adopt a $B-V$ colour cut of 0.8 at $z = 0.0$, and use $B-V$ values of 0.70 and 0.65 as colour cuts at \mbox{$z = 0.5$} and $z = 1.0$ respectively. 
We have found that 67\%, 52\%, 37\% of our galaxy-sized haloes (and 97\%, 97\%,  94\% of our group-sized haloes) are associated with red central galaxies at \mbox{$z = 0.0$, 0.5 and 1.0} respectively.

For nearly all of the observed lensed systems, with which we wish to compare, the projected (physical) separation between the main lensing galaxy and the luminous satellite galaxy is about 1 arcsecond, corresponding to 4.2 $h^{-1}$ kpc at \mbox{$z = 0.5$} and 5.5 $h^{-1}$ kpc at $z = 1$.  To explore whether current simulations produce enough satellite galaxies in the inner region of a galaxy halo to explain flux anomalies, we have counted all satellite galaxies (satisfying the restrictions outlined above) within a 5 $h^{-1}$ kpc projected region from the centre of the host galaxy (defined as the position of the most bound particle), and will refer to these galaxies as bright central substructures.  We also explore the effect of increasing this to a 10 $h^{-1}$ kpc projected region. 

\section{Results}

\label{results}

\subsection{Theoretical predictions}
\label{sec:predictions}

\begin{figure*}
\begin{center}
\begin{tabular}{lll}

\includegraphics[width=6cm,height=6cm,angle=-90,keepaspectratio]{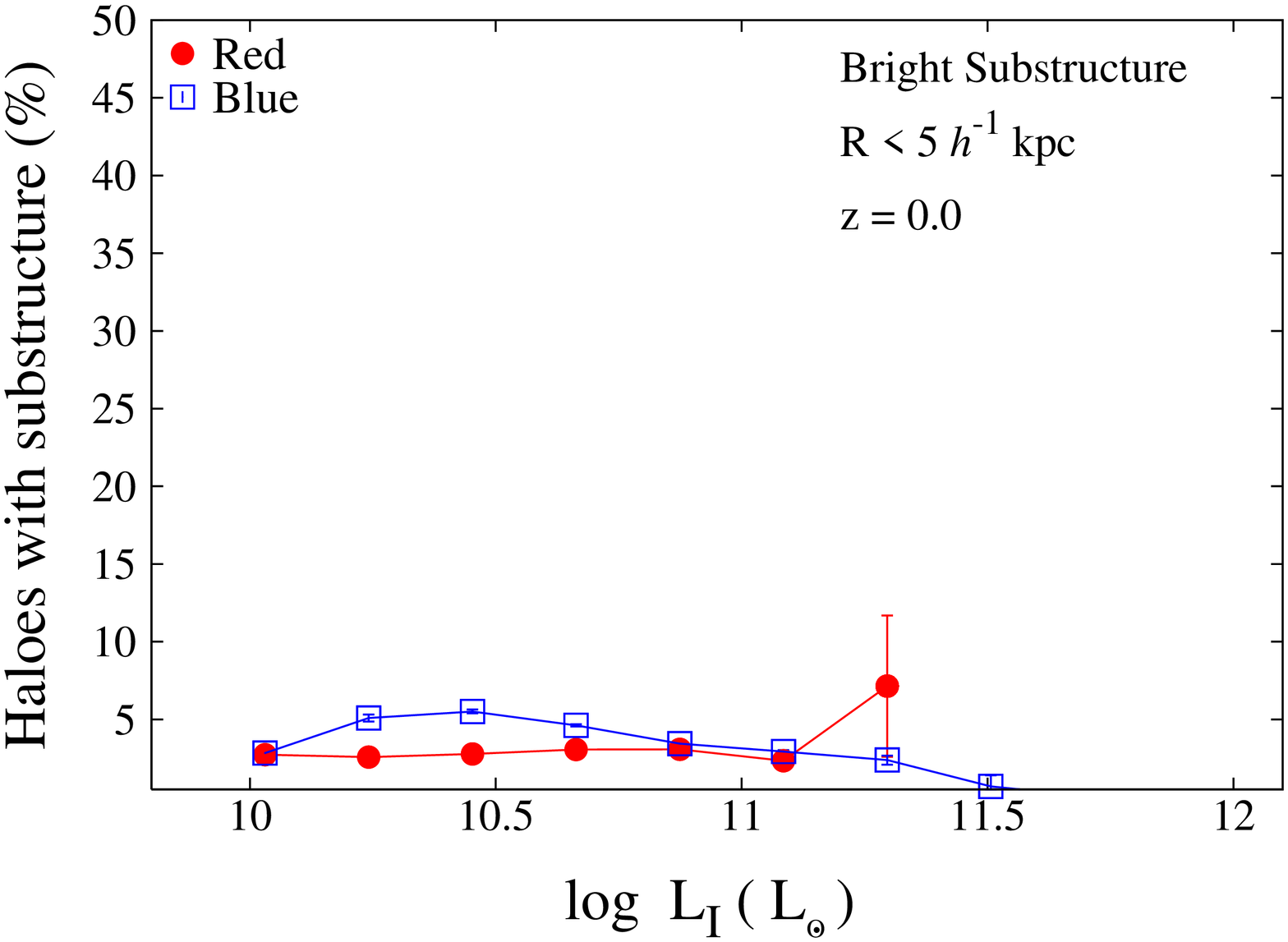} & 
\hspace{-0.4cm}\includegraphics[width=6cm,height=6cm,angle=-90,keepaspectratio]{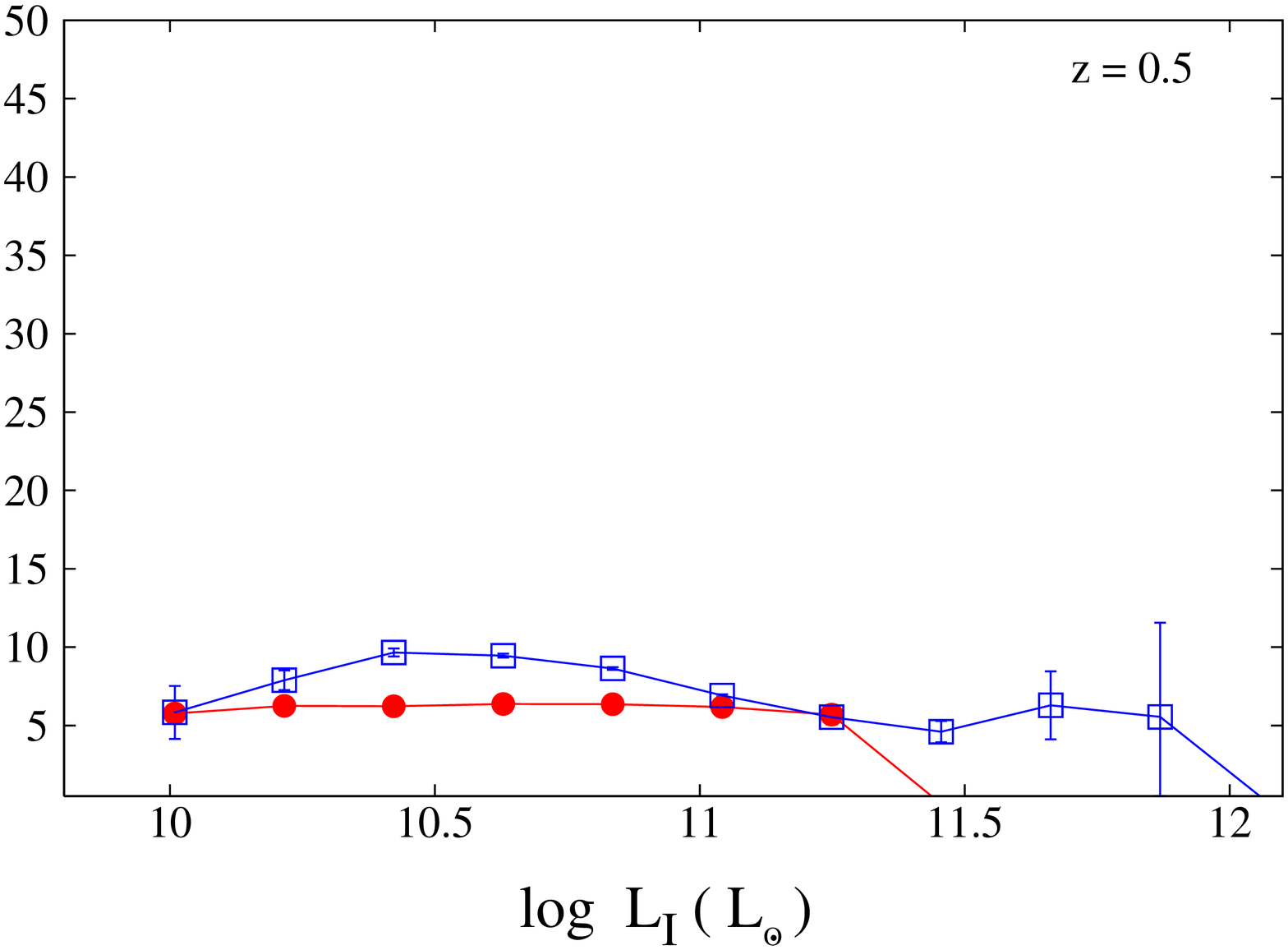} &
\hspace{-0.4cm}\includegraphics[width=6cm,height=6cm,angle=-90,keepaspectratio]{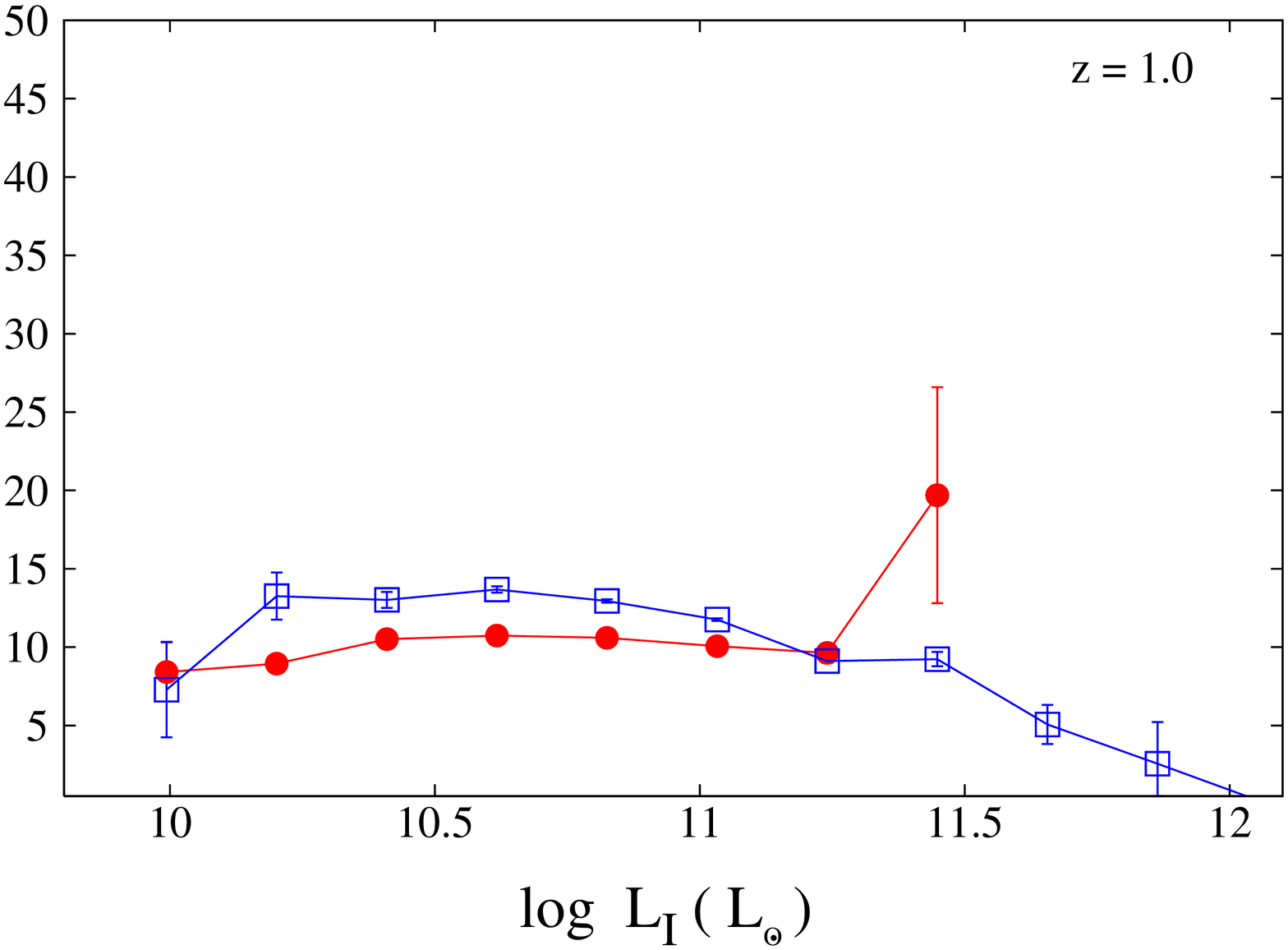} \\

\includegraphics[width=6cm,height=6cm,angle=-90,keepaspectratio]{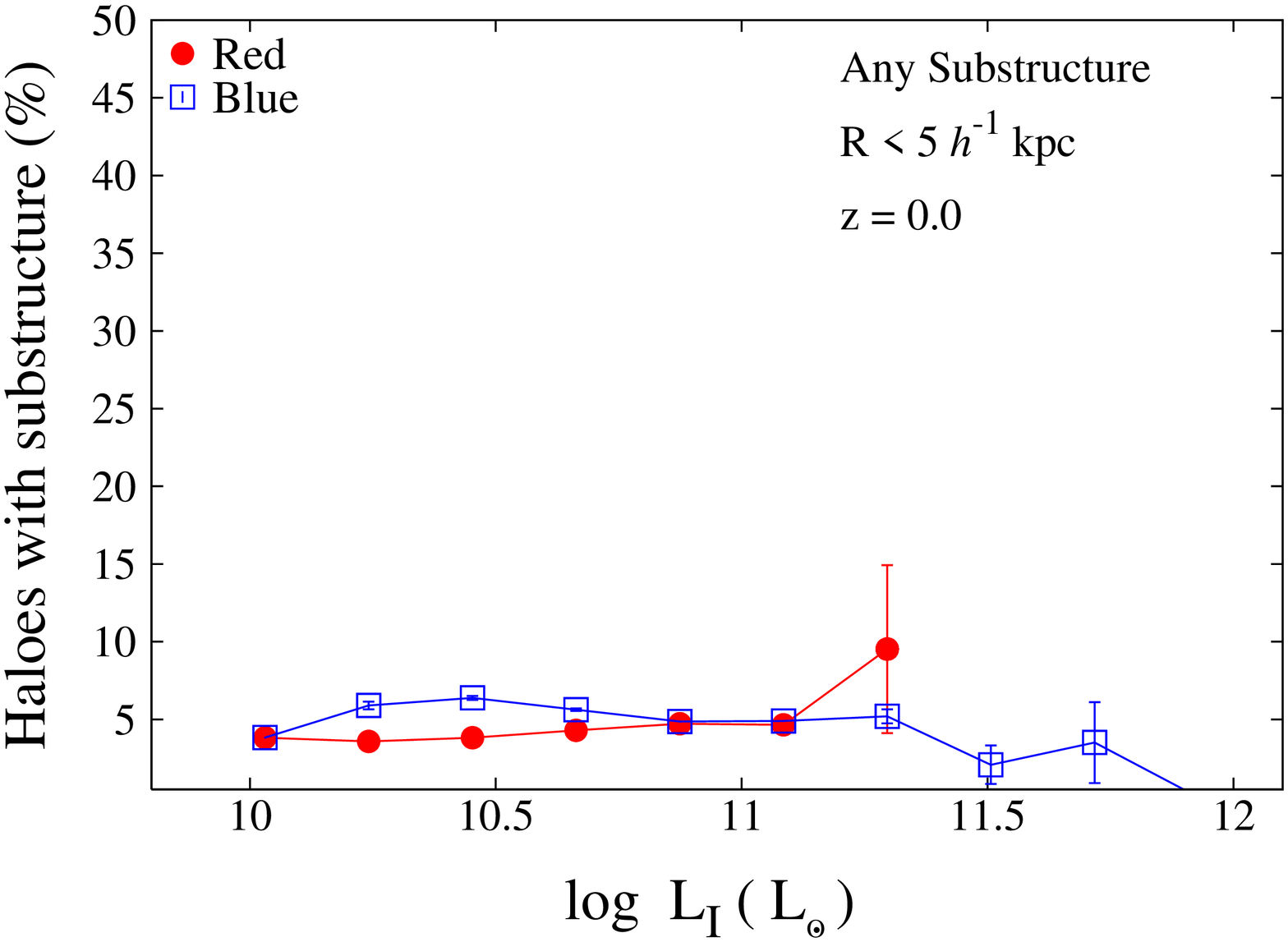}  &
\hspace{-0.4cm}\includegraphics[width=6cm,height=6cm,angle=-90,keepaspectratio]{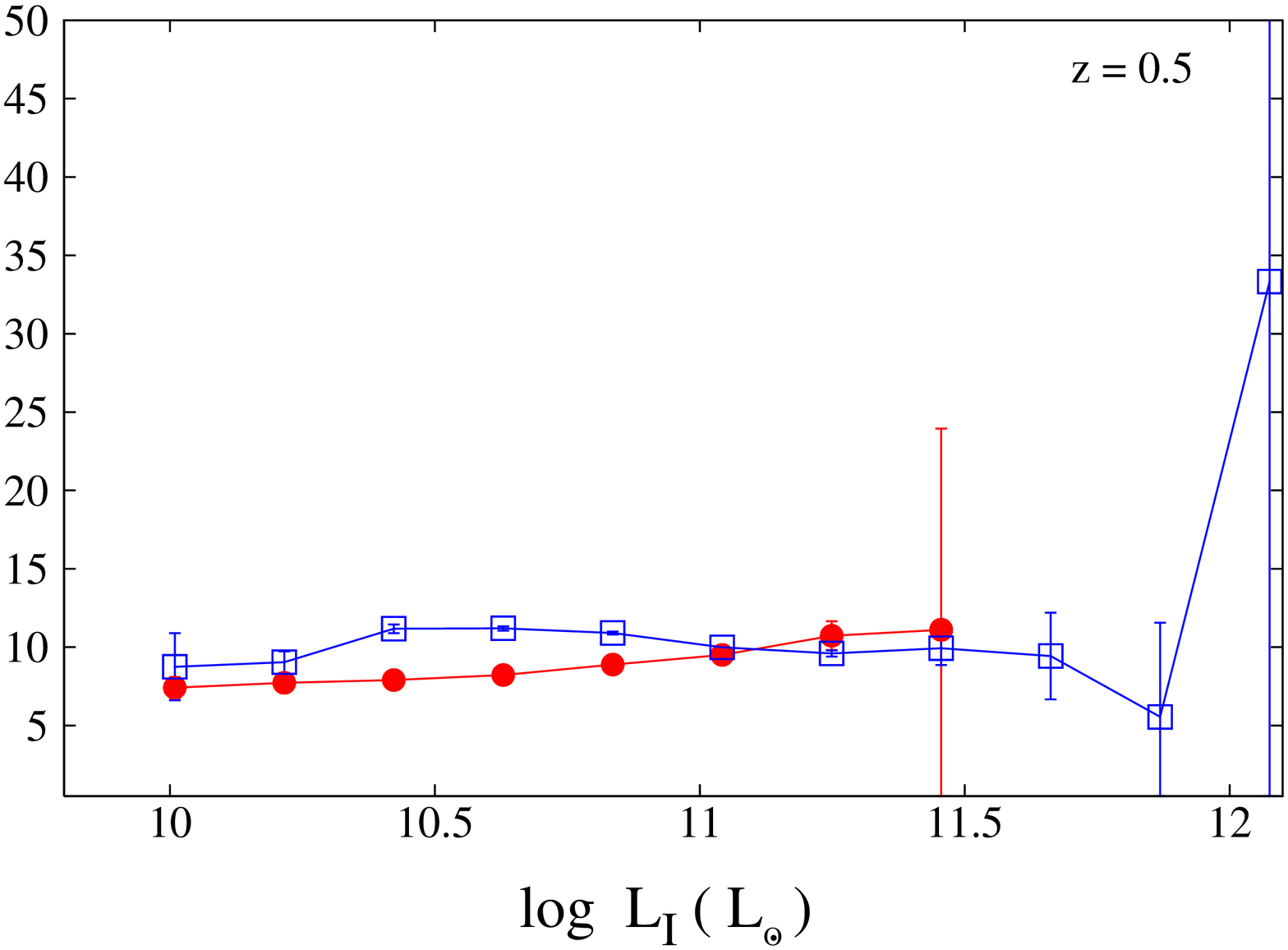}  &
\hspace{-0.4cm}\includegraphics[width=6cm,height=6cm,angle=-90,keepaspectratio]{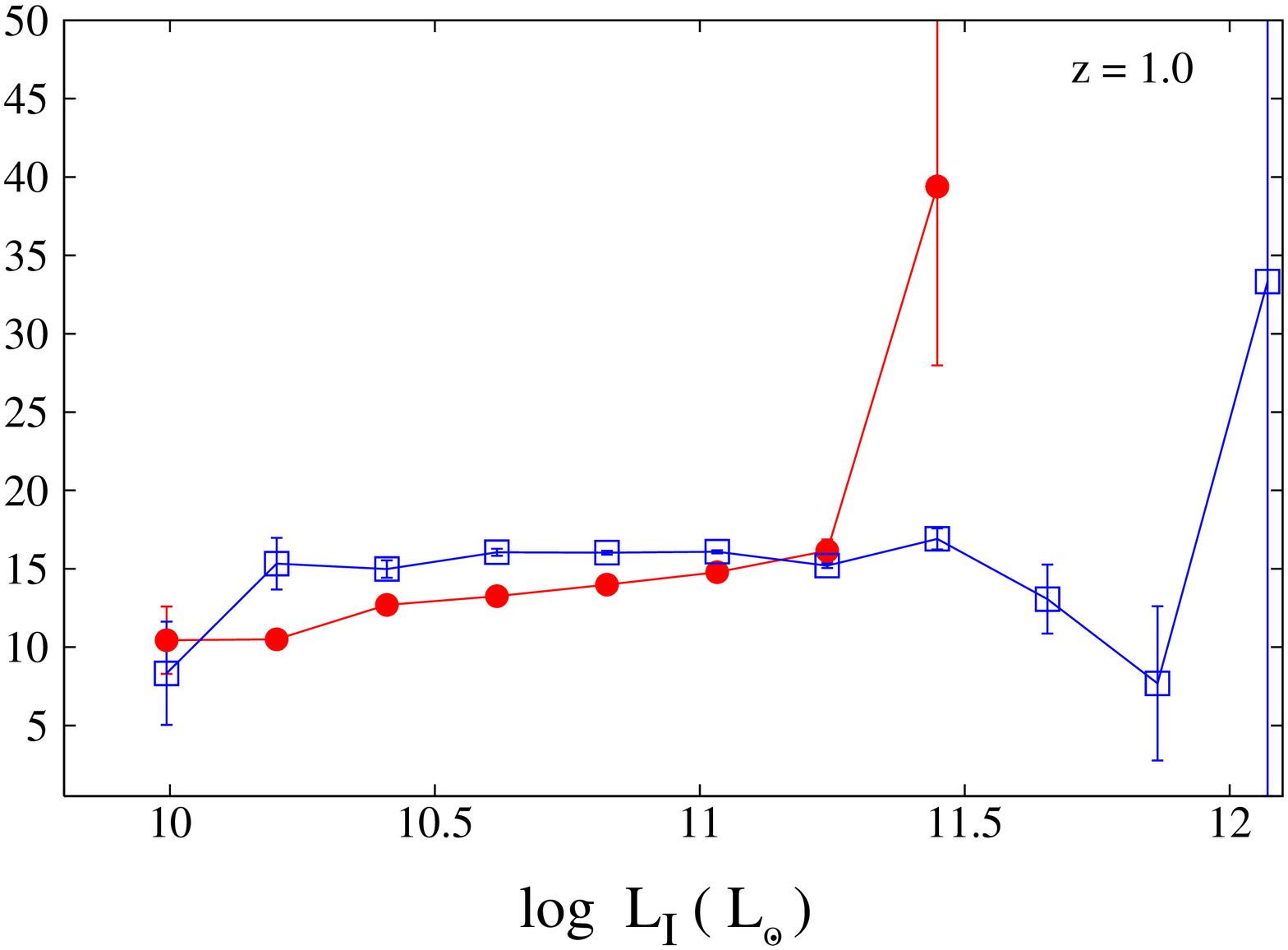} 

\end{tabular}
\end{center}
\caption{\label{within}The percentage of galaxy-sized haloes ($10^{12} h^{-1}$ M$_\odot < \mbox{M}_{vir} < 10^{13} h^{-1}$ M$_\odot$) which have substructure within the inner region of the halo, as a function of luminosity.  The top row shows the fraction of haloes with bright satellites within the central 5 $h^{-1}$ kpc (projected).  The bottom row shows the fraction of haloes which contain
any substructure within the central 5 $h^{-1}$ kpc (projected) region.  In these plots the red population is depicted using filled circles, while
the blue population is shown using squares.  The three columns show the fraction of haloes containing substructure for the
three different redshifts we have considered (left: 0.0, middle: 0.5 and right: 1.0).  The Poisson scatter is shown.  Note that the most luminous blue hosts are not necessarily the most massive haloes but are likely to have undergone recent star formation.
}
\end{figure*}

In Fig. \ref{within}, we show the
fraction of galaxy-sized (\mbox{$10^{12} \, h^{-1}$ M$_\odot < $} \mbox{M}$_{vir} < 10^{13} \,h^{-1}$ M$_\odot$) haloes with satellite galaxies, satisfying our magnitude cuts, within the central region
(projected) of the halo.  The fraction is plotted as a function of the host galaxy's I-band luminosity (for ease of comparison with the CLASS data).  The fraction shown is the mean value, when averaged over 3 independent projections.  The uncertainty corresponds to the Poisson scatter within each bin.
The top row shows the fraction of haloes with bright satellites within the central 5 $h^{-1}$ kpc (projected) while the bottom row shows the fraction of haloes which contain any (dark or bright) substructure within the same region.  In these plots the red population is depicted using filled circles, while
the blue population is shown using squares.  The three columns show the fraction of haloes containing substructure for the
three different redshifts we have considered (left: 0.0, middle: 0.5 and right: 1.0).  Note that the most luminous blue hosts are not necessarily the most massive haloes but are likely to have undergone recent star formation.

We find that at $z = 0$, about $3 - 4$\% of all of our galaxy-sized haloes appear to have bright satellite galaxies within \mbox{5 $h^{-1}$ kpc} (projected) of the centre of the host.  While this fraction is similar in both types of galaxies, it is found to increase with redshift (rising to around \mbox{ 11 $ - $ 12\%} at \mbox{$z = 1$}).  Extending the central region to \mbox{10 $h^{-1}$ kpc} (not shown) we find that about 10\% of our galaxy-sized haloes host bright central substructure at $z = 0$, and that this increases to almost \mbox{27\% at $z = 1$.} 
If we consider all substructure within the projected central region, not restricting the search to `observable' satellite galaxies, the fractions do not change significantly.  This is illustrated in the bottom panels, where we have dropped our lower magnitude and luminosity ratio cuts.  The fraction of haloes containing any substructure within \mbox{5 $h^{-1}$ kpc} (projected) of the centre of the host  increases only moderately from $\sim$ 3\% to about 5\% at $z = 0$.  This conclusion remains valid all the way up to $z = 1$, where the fraction increases from $\sim$ 11\% to about 15\%.  

\begin{figure*}
\begin{center}
\begin{tabular}{lll}

\includegraphics[width=6cm,height=6cm,angle=-90,keepaspectratio]{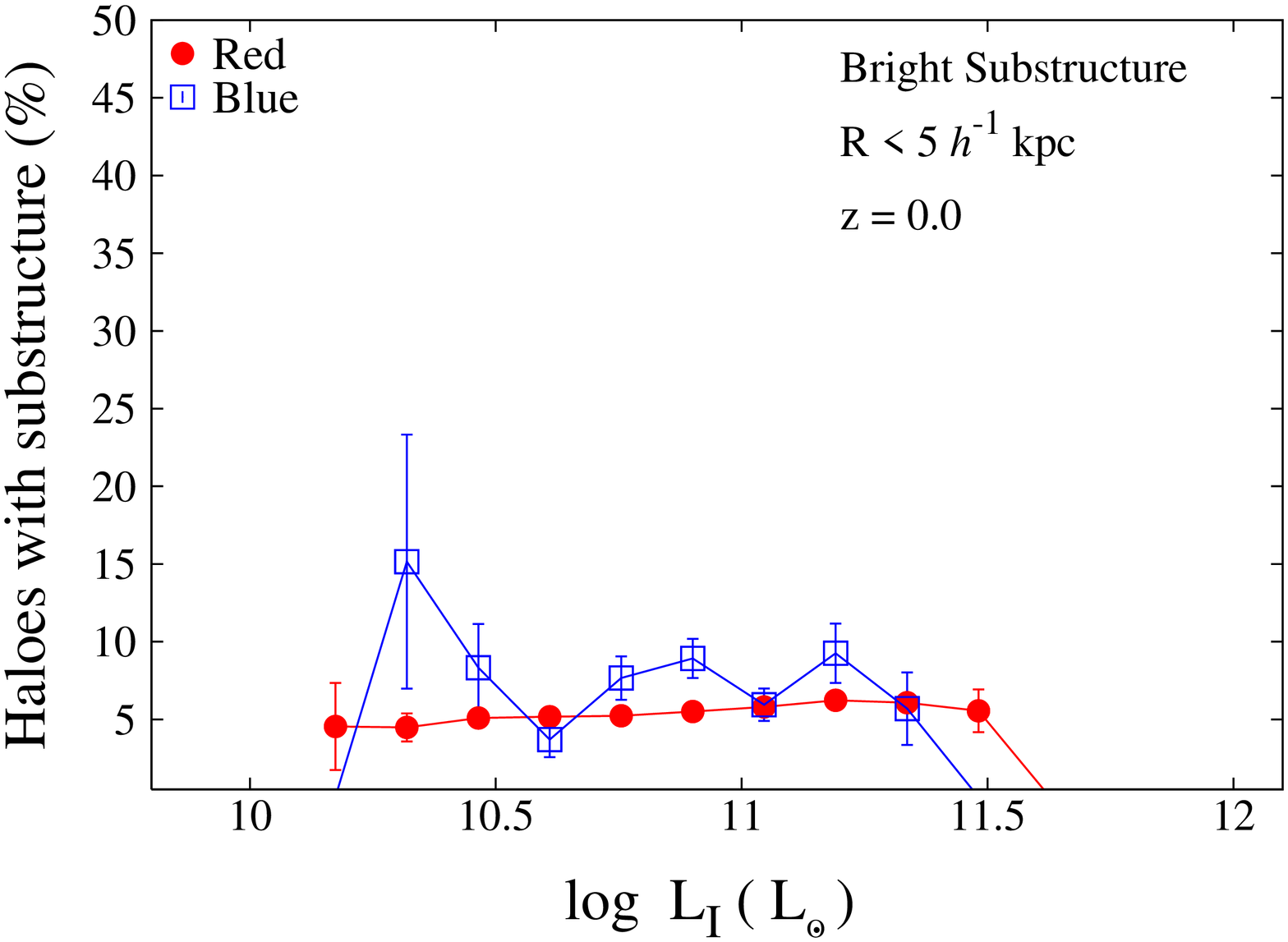} & 
\hspace{-0.4cm}\includegraphics[width=6cm,height=6cm,angle=-90,keepaspectratio]{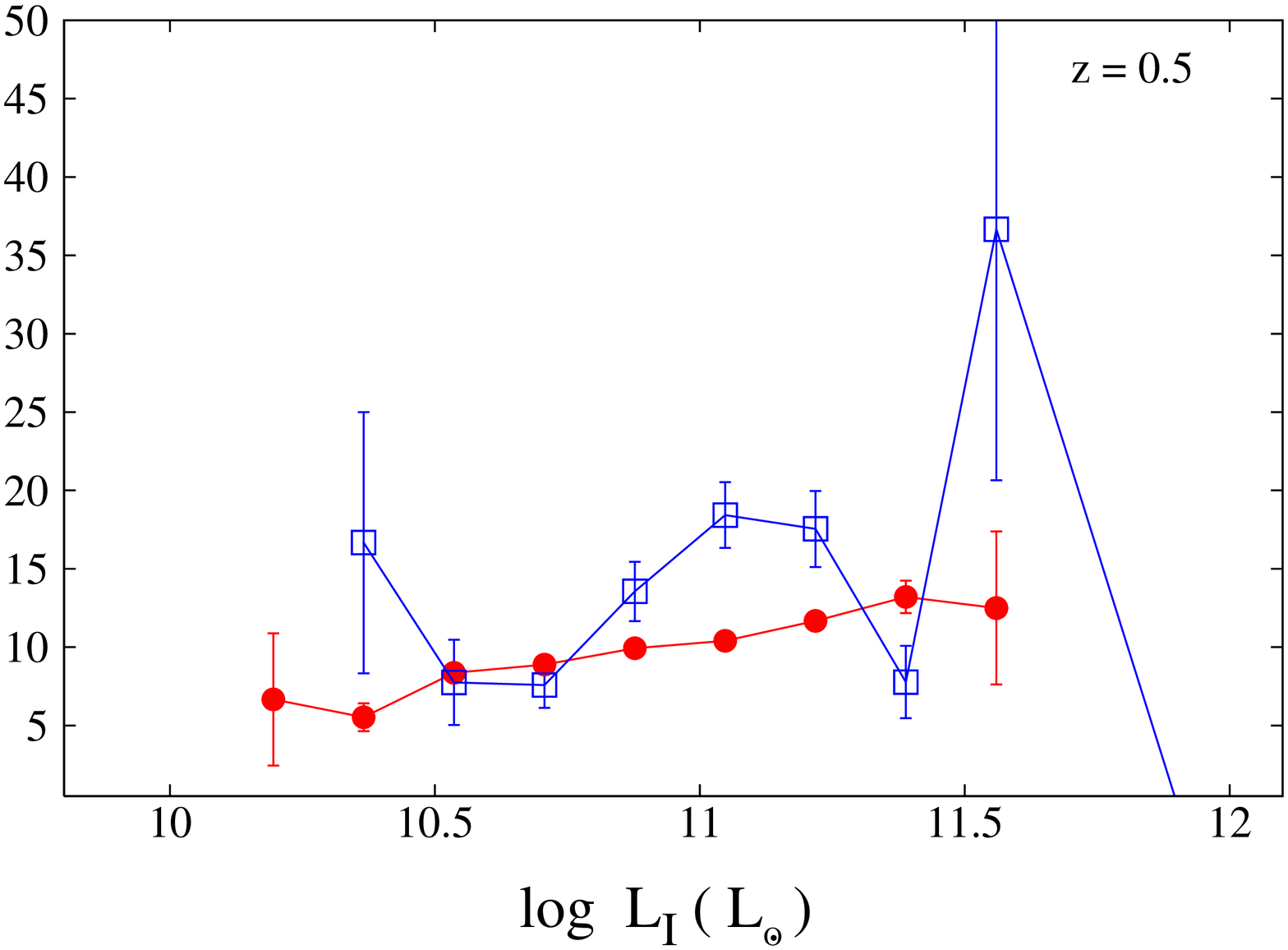} &
\hspace{-0.4cm}\includegraphics[width=6cm,height=6cm,angle=-90,keepaspectratio]{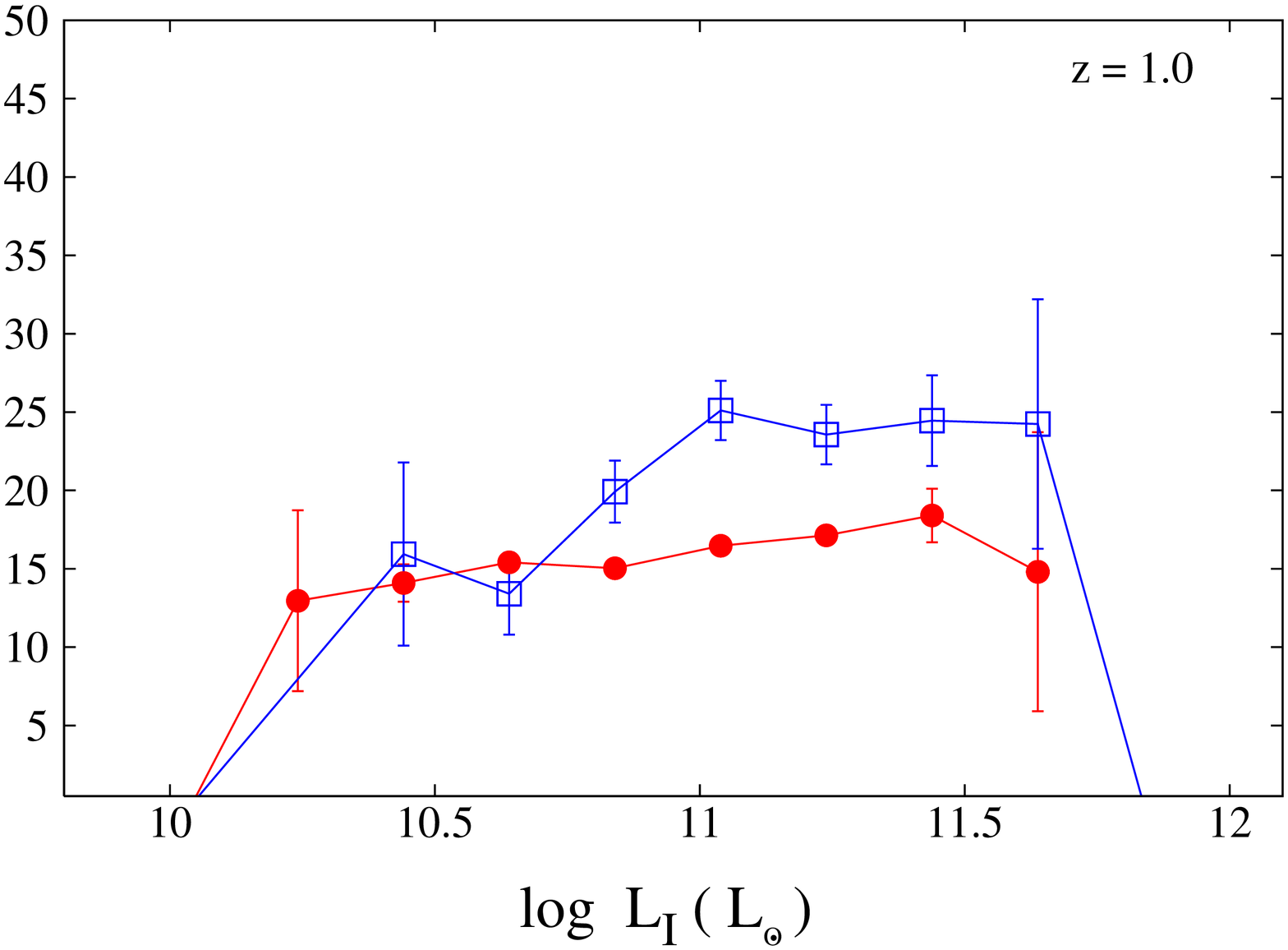} \\

\includegraphics[width=6cm,height=6cm,angle=-90,keepaspectratio]{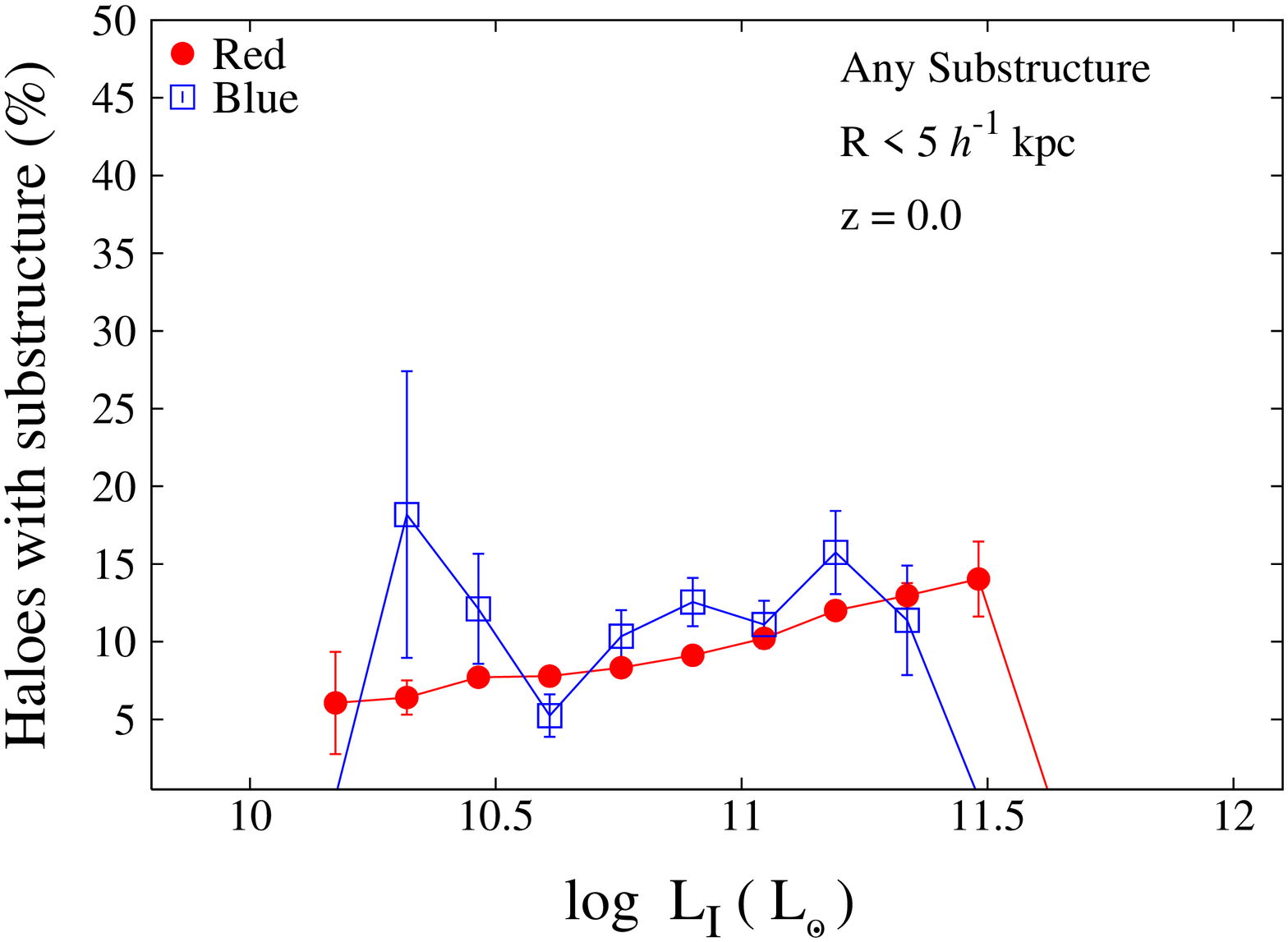}  &
\hspace{-0.4cm}\includegraphics[width=6cm,height=6cm,angle=-90,keepaspectratio]{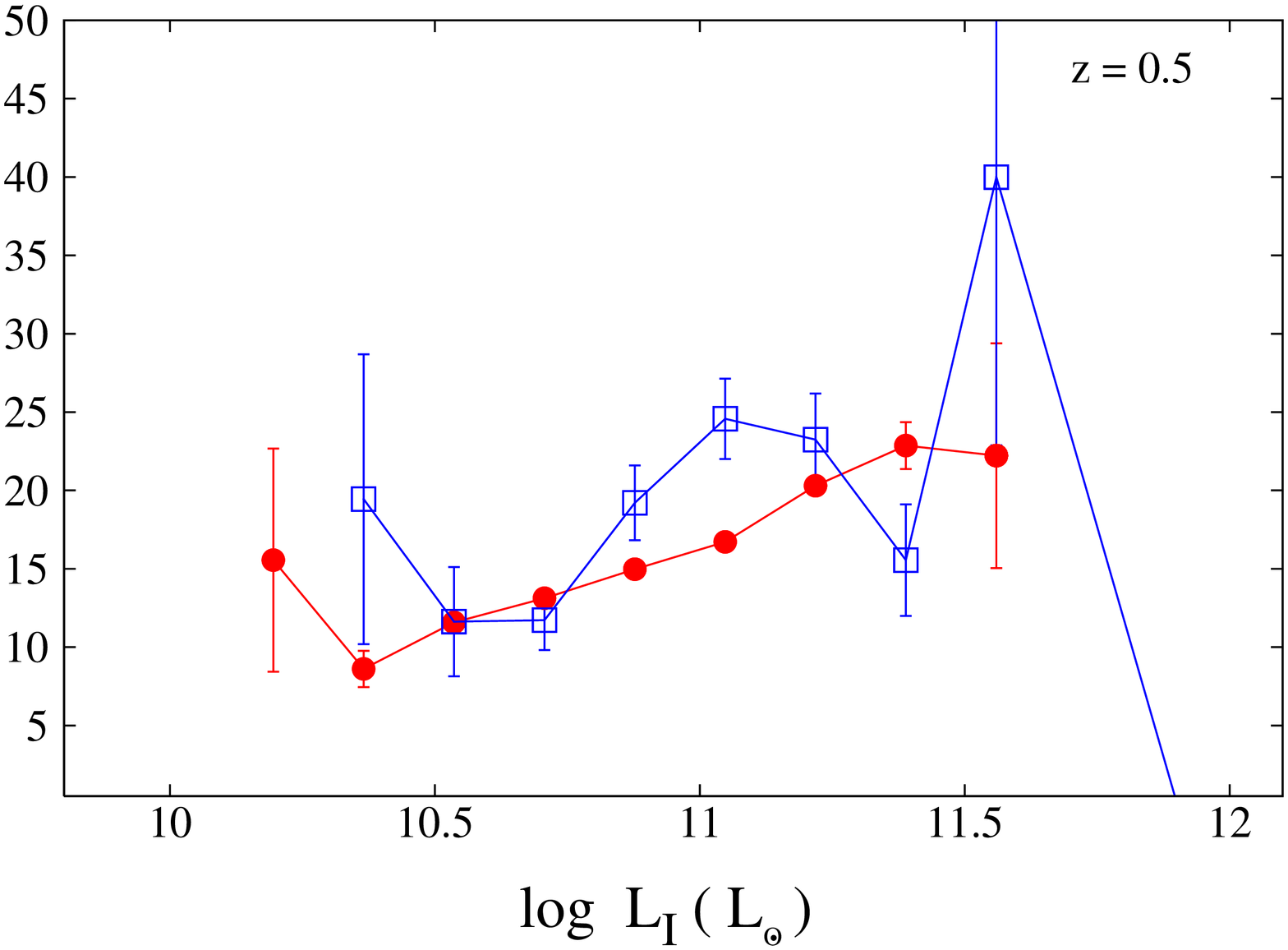}  &
\hspace{-0.4cm}\includegraphics[width=6cm,height=6cm,angle=-90,keepaspectratio]{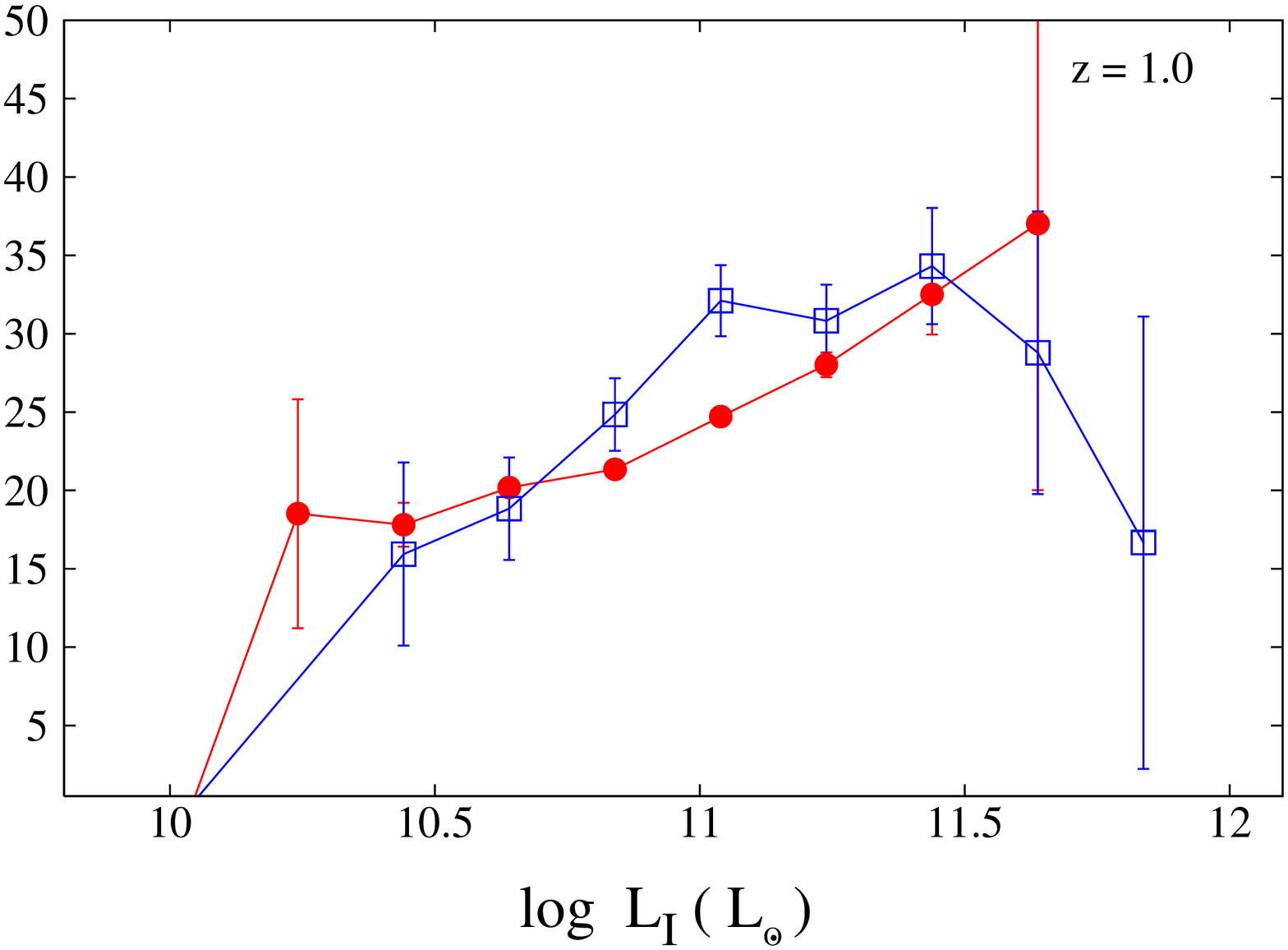} 

\end{tabular}
\end{center}
\caption{\label{fig:group} As in Fig. \ref{within} but for group-sized haloes ($10^{13} h^{-1}$ M$_\odot \leq \mbox{M}_{vir} < 10^{14} h^{-1}$ M$_\odot$).  The top row shows the fraction of haloes with bright satellites within the central 5 $h^{-1}$ kpc (projected).  The bottom row shows the fraction of haloes which contain
any substructure within the central 5 $h^{-1}$ kpc (projected) region.  In these plots the red population is depicted using filled circles, while
the blue population is shown using squares.  The three columns show the fraction of haloes containing substructure for the
three different redshifts we have considered (left: 0.0, middle: 0.5 and right: 1.0).  The Poisson scatter is shown.}
\end{figure*}

Since some of the lensing galaxies reside in groups (see Section \ref{sec:observations}), we explicitly check the fraction of haloes with bright satellites within a projected 5 $h^{-1}$ kpc region in group-sized haloes (with \mbox{$10^{13} h^{-1}$ M$_\odot  \leq M_{vir} < 10^{14} h^{-1}$ M$_\odot$}).  The results are shown in Fig. \ref{fig:group}. The fraction of group-sized
haloes with bright central substructure is higher than in galaxy-sized haloes, increasing to $\sim 6\%$ at \mbox{$z = 0$} and to $\sim$ 16\% at $z = 1$ for red galaxies, which dominate the lensing cross-sections; the fraction is slightly higher for blue galaxies.

The lensing cross-section is roughly proportional to $\sigma^4$ (e.g. \citealt{1984ApJ...284....1T}),
 where $\sigma$ is the velocity dispersion of the system
 and, from the Faber-Jackson relation,
$L \propto \sigma^4$ for ellipticals (\citealt{bib:Faber76}), thus to compare with observations,
the fraction should be weighted by luminosity. The luminosity weighted fraction of hosts with bright central substructure is
given in Table \ref{fractions}.  At \mbox{$z = 0$}, the fraction is of the order of $3\%$ for galaxy-sized haloes, increasing to about $6\%$ if we consider only group-sized haloes.  At $z = 1$, the fraction for galaxy-sized haloes is about 11\%, rising to $\sim$ 17\% for group-sized haloes, still slightly below (but possibly consistent with) the observed fraction of galaxies
with bright companions (see Section \ref{sec:observations}).

Of the systems found to host bright central substructure, most have only one bright central satellite.  Only 2\% (3\%) of galaxy-(group-) sized hosts with bright central substructure host more than one bright central satellite at $z = 0$.  At $z = 1$, 5\% (7\%) of galaxy-(group-) sized hosts have multiple bright central satellites.  The largest number of bright central satellites found within any one system is 4.

The force softening of the simulation is 5 $h^{-1}$ kpc (in comoving coordinates); within this region, resolution effects may be significant.  For this reason, we explicitly check the fraction of the projected subhaloes within the 3D central region.  The fraction depends strongly on redshift.  About 32\% of the luminous satellites found within the projected central 5 $h^{-1}$ kpc are found within the 3D central region in galaxy-sized haloes at $z = 0$ (for groups this decreases to $\sim $ 24\%). At $z = 1$ this fraction is 11\%, dropping to $\sim$ 6\% in group-sized haloes (see Table \ref{3d}).  All of the satellites within the 3D central region are `orphan' galaxies (see Section \ref{discussion}).  

\begin{table}
\caption{ \label{fractions}Luminosity weighted fraction of galaxy-sized hosts with bright substructure within the central 5 $h^{-1}$ kpc (projected).  Numbers in brackets correspond to values for group-sized haloes.}
\centering
\begin{tabular}{@{}|l||ccc|@{}}
\hline
 & \multicolumn{3}{c}{Redshift} \\
 & 0.0 & 0.5 & 1.0 \\
\hline 
All hosts  & 3 (6)$\%$ & 7 (11)$\%$ & 11 (17)$\%$\\
Red hosts  & 3 (6)$\%$ & 6 (11)$\%$ & 11 (16)$\%$   \\
Blue hosts & 4 (7)$\%$ & 7 (16)$\%$ & 11 (24)$\%$\\
\hline
\end{tabular}
\end{table}

\begin{figure*}
\begin{center}
\includegraphics[width=8.5cm,height=8.5cm,angle=-90,keepaspectratio]{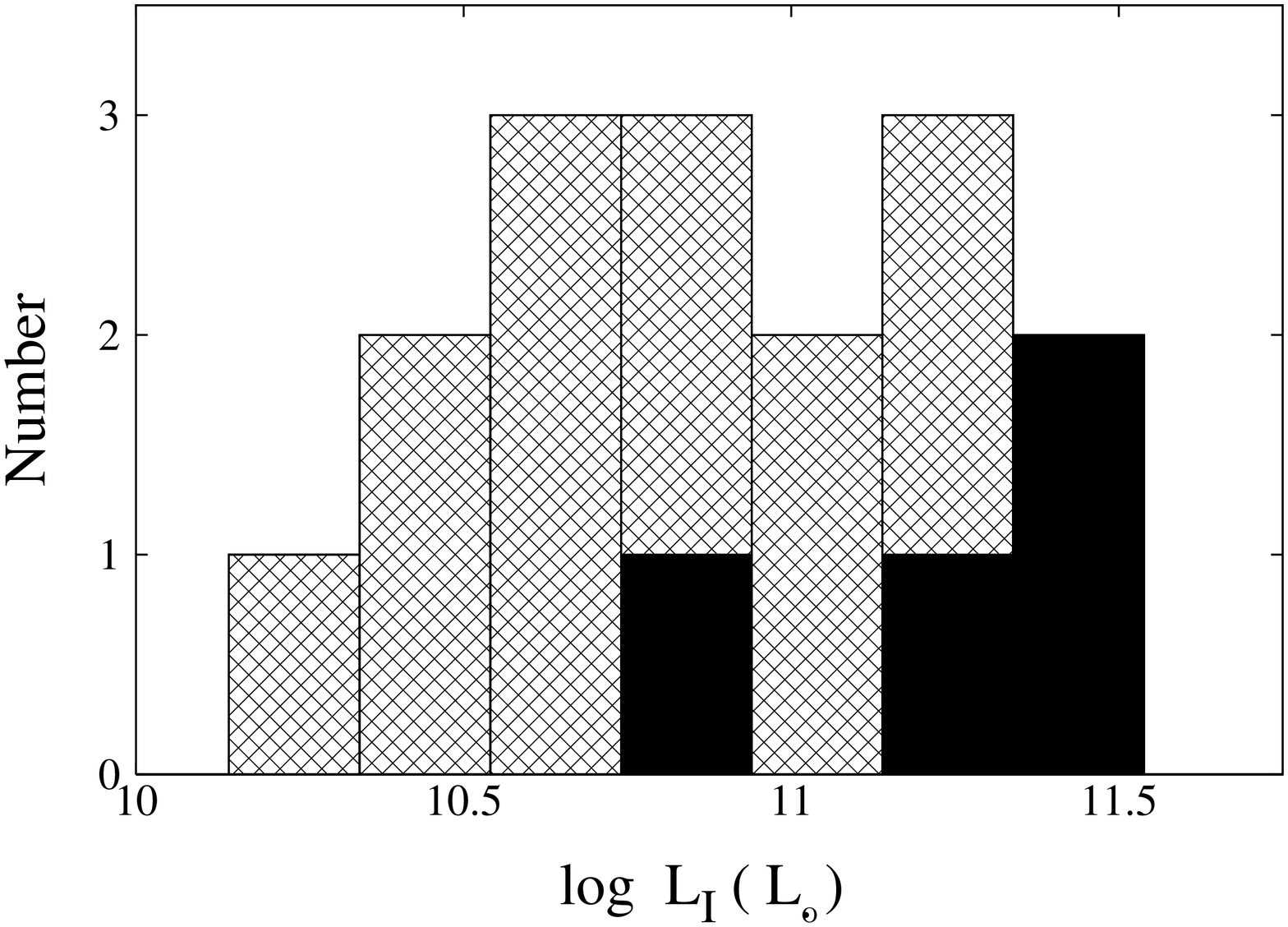}  
\includegraphics[width=8.5cm,height=8.5cm,angle=-90,keepaspectratio]{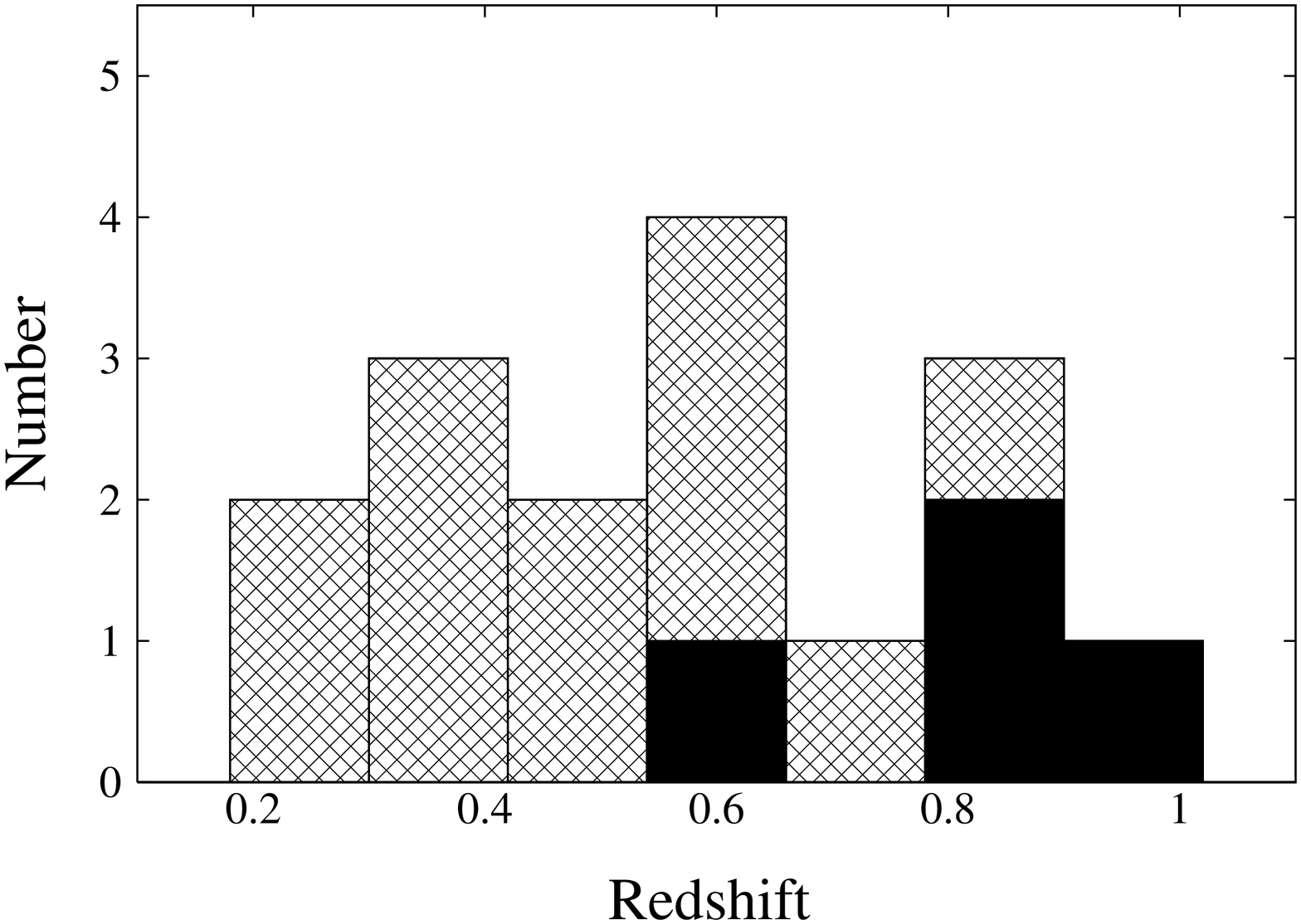} 
\caption{\label{obs} Histogram showing the distribution of the 16 CLASS lenses with available redshifts and I-band magnitudes, while the solid histogram shows the distribution of the 4 CLASS lenses with luminous satellites (B1127+385, is not shown due to its uncertain lens redshift).  K-correction values have been taken from \citet{bib:Poggianti97}, interpolated using a polynomial fit.}
\end{center}
\end{figure*}

\begin{table}
 \caption{\label{3d}Fraction of projected central satellites within the 3D central region of galaxy-sized hosts.  Numbers in brackets correspond to values for group-sized haloes.}
\centering
\begin{tabular}{@{}|l||cc|@{}}
 \hline
  &  $r_{3d} < $ force softening  \\
  \hline
  z = 0.0 Bright & 32 (24) \%  \\
  z = 0.0 Total & 26 (17) \%  \\
  \hline
  z = 0.5 Bright & 17 (11) \% \\
  z = 0.5 Total & 14 (8)\% \\ 
  \hline
  z = 1.0 Bright & 11 (6)\% \\
  z = 1.0 Total & 9 (5) \%  \\
  \hline
\end{tabular}
\end{table}

\subsection{Substructure in CLASS lenses}
\label{sec:observations}

We use the CLASS survey as the primary observational data. This survey discovered 22 new gravitational lenses in the radio (\citealt{bib:Browne03};
\citealt{bib:Myers03}). \citet{bib:Kochanek04} used 6 lenses from this survey to study the amount of substructure in lensing galaxies. In this work, we have used the whole survey to gather statistics; 5 of the 22 CLASS lenses have luminous satellite galaxies within \mbox{5 $ h^{-1}$ kpc} of the main lensing galaxy:  B1608+656, B2045+265, MG\,0414+0534, B1127+385 and B1359+154.

For B1608+656, the main lensing galaxy (G1) is at redshift $z_l=0.63$. In addition, there is a faint galaxy, G2, about 0.73 arcsec away, which is 1.8 magnitudes fainter than G1 both in the HST F160W and F814W (\citealt{bib:Koopmans03}) filters.  There are also four groups along the line of sight, including one at the redshift of the lensing galaxy G1 (\citealt{bib:Fassnacht06}).  If G2 is at the same redshift as G1, then the projected separation is \mbox{3.4 $ h^{-1}$ kpc}.

For B2045+265, \cite{bib:McKean07} found a galaxy, G2, about 0.66 arcsec
away from the main lensing galaxy G1 (at redshift 0.867), which is
between 3.6 and 4.5 magnitudes fainter than the main lensing galaxy G1 in
F814W and F160W. The photometric redshift of G2 is consistent with that of G1, although it is also consistent with being at redshift $\sim 4-5$.
If the redshifts of G1 and G2 are the same, then the projected
separation is 3.5 $ h^{-1}$ kpc.

For the quadrupole lens MG\,0414+0534, \cite{bib:Schechter93} found a fainter companion that is about 1 arcsec away from the main lensing galaxy (which is at redshift 0.96). The object `X' is about $2.6 - 2.44$ magnitudes fainter than the main lensing galaxy in the HST images of F160W and F814W. If the object `X' is at redshift 0.96, then the projected separation is 5.5 $ h^{-1}$ kpc.

For B1127+385, there are also two lensing galaxies, G1 and G2 (\citealt{bib:Koopmans99}). The fainter one, G2, is about 1 magnitude fainter than G1 in both F814 ($I=22.5$ for G1) and F555 ($V=24.4$ for G1). The separation between these two galaxies is about 0.6 arcsec. If the lensing galaxy's redshift is between 0.5 and 1, then the projected separation is about $2.5 - 3.3$ $ h^{-1}$ kpc.

B1359+154 is a six image system produced by a small group of
galaxies. Three primary lens galaxies lie on the vertices of a
triangle separated by 0.7 arcsec at z $\sim 1$ (corresponding to a
projected separation of $\sim$ 3.9 $ h^{-1}$ kpc), with magnitudes in $I$ of 
$22.68\pm 0.28$ and $23.69\pm 0.24$ and $23.70\pm 0.33$ (\citealt{bib:Rusin01}).

In the left panel of Fig. \ref{obs} we show the luminosities of the observed lenses found by the CLASS survey.  The patterned histogram shows 16 of the CLASS lenses for which we have redshifts and I-band magnitudes (taken from the CASTLES website\footnote{http://www.cfa.harvard.edu/castles/ (\citeauthor{bib:Castles})}).  Over-plotted (solid histogram) are the 4 CLASS lenses which have been shown to host luminous satellites; for B1127+385, its luminosity is unknown due to the uncertain lens redshift.

The right panel of Fig. \ref{obs} shows the redshift distribution of the CLASS lenses.  All of the lenses with luminous satellites have redshifts higher than the median value of \mbox{$\sim$ 0.6.} About 75\% of the lenses with z $>$ 0.8 have luminous satellites.  We caution that the six remaining lenses of the CLASS sample (with unknown redshifts) may be, on average, at higher z.  By ignoring these lenses, the redshift distribution may be somewhat skewed.  This could mean that the probability of a high redshift lens hosting a luminous satellite may not be as high as implied.

\begin{figure}
\begin{center}
\includegraphics[width=9.5cm,height=9.5cm,angle=-90,keepaspectratio]{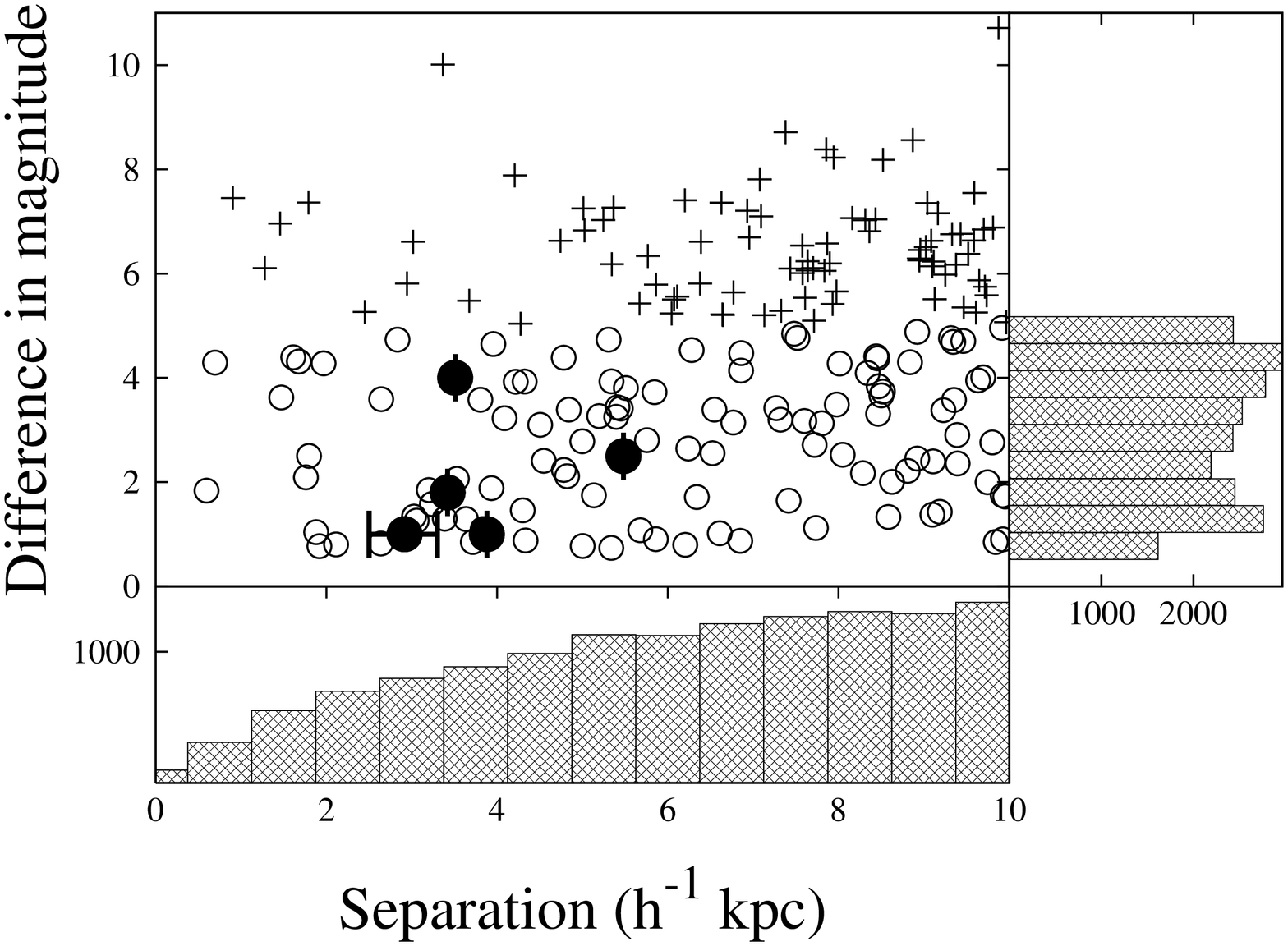} 
\end{center}
\caption{\label{deltam} Difference in I-band magnitude between the host and satellite galaxy versus the projected separation (in $h^{-1}$ kpc) of the satellite galaxy from the host. We show a random selection of our group-sized haloes (selected at redshift 1.0) with `dark' substructure (crosses) and  bright substructure (circles).  The 5 CLASS lenses found to have luminous satellite galaxies are plotted with solid circles; for B1127+385, the horizontal bar shows the range of separations when the lens redshift is varied from 0.5 to 1.  The histograms show the distribution of bright substructure found within the central 10 $h^{-1}$ kpc (projected) as a function of separation and magnitude difference.}
\end{figure}

Fig. \ref{deltam} shows the difference in magnitude between the host and satellite galaxy versus the projected separation of the satellite galaxy from the host. We show a random selection of our group-sized haloes with `dark' substructure (crosses) and  bright substructure (circles) from our $z = 1$ sample.  The 5 CLASS lenses found to have luminous satellite galaxies are plotted with solid circles; for B1127+385, the horizontal bar shows the range of separations when the lens redshift is varied from 0.5 to 1. Selection effects may be complicated and have not been taken into account in this study.  While it will be difficult to observe satellites with large magnitude differences at small separations, we find that there are also few simulated satellites at very small separations. (The increase in number with separation is due to the larger area considered).  As illustrated with the histograms in Fig. \ref{deltam} we find that our sample of host galaxies and their luminous satellites is comparable to the observed galaxies in the (small) CLASS sample. 

\subsection{Resolution effects}
\label{discussion}
\begin{table*}
\caption{\label{types}Percentage of bright satellite galaxies without a surviving dark matter subhalo within the virial radius and within the central \mbox{5 $h^{-1}$ kpc} (projected) region for galaxy-sized hosts. Numbers in brackets correspond to values for group-sized haloes.}

\centering
\begin{tabular}{@{}|l||ccc|@{}}
\hline
 & \multicolumn{3}{c}{Redshift} \\
 & 0.0 & 0.5 & 1.0 \\
\hline 
Satellites without DM subhalo & 74 (68)\%& 81 (76)$\%$ & 87 (83)$\%$  \\
\hline
Projected central satellites without DM subhalo & 98 (98)$\%$ & 99 (98)$\%$ & 99 (99)$\%$ \\
\hline
\end{tabular}
\end{table*}

As haloes fall into a larger system, they are exposed to tidal forces and are stripped as they orbit the host system.  The extent to which a halo is stripped depends on resolution and the inner density profile of the halo (\citealt{bib:Moore96}).  The simulated subhaloes have artificially low density cores (due to force softening) that makes them more susceptible to tidal stripping.  Including baryons (and gas cooling) will increase the central density and make the
galaxy more resistant to tidal stripping (\citealt{bib:Moore96, bib:Maccio06}), although the cooling of baryons toward the central host will also increase the tidal forces experienced by subhaloes that come close to the centre.  

Nearly all of the satellite galaxies we find in the projected central regions are tidally stripped `orphan' galaxies (see Table \ref{types}).  The semi-analytic model we have used follows the orbit of galaxies which have lost their dark matter subhalo, by assuming that they follow the motion of the most bound particle of the parent subhalo before it was destroyed (this is shown to be a good estimate of the subhalo's position by \citealt{bib:Springel01}).  Since the effects of dynamical friction on the orbit of these galaxies is not considered in detail, caution is required when interpreting our results.   As noted in \cite{bib:Sales07}, this may affect the overall abundance and radial distribution of these stripped haloes.  

Also, it is assumed that the `orphan' galaxy remains completely undisturbed for a merging time, based on the dynamical friction formula of \cite{bib:Binney87}, until it merges with the central galaxy.  This assumption may result in an over-estimate of the number of `orphan' galaxies and their associated luminosity.  \cite{bib:Henriques07} take the opposite approach and assume that all `orphan' galaxies which have not merged with the central galaxy by $z = 0$ are completely disrupted, and are responsible for the diffuse intracluster light.  While their results suggest an improved match to the luminosity function in groups and clusters, the model is simplistic.  They note that it is more likely that the disruption would happen gradually, and that the dense cores may survive for longer.  The extent to which disruption would affect these `orphan' galaxies remains unclear. 
However, survival of these `orphan' galaxies (at least to some extent) has been shown by \cite{bib:Wang06} to be essential in order to explain the observed correlation signal at small scales.  This study used the Millennium Simulation to construct a new model of galaxy clustering. They found that if `orphan' galaxies were excluded from the analysis, the correlation signal decreases at small scales in contrast to observations.

 A related question is: if we were to increase the numerical 
resolution of the simulation, would the fraction of luminous satellites rise significantly? Clearly, the number of subhaloes (dark or luminous) must rise further 
since the subhalo mass function roughly follows a power law with $dn/dM \propto M^{-\alpha}$, $\alpha = 1.7 - 1.9$ (\citealt{bib:Moore99}; \citealt{bib:Ghinga00}; \citealt{bib:DeLucia04}; \citealt{bib:Gao04_dm}; \citealt{bib:Diemand07}).  The lowest mass subhaloes we can resolve have circular velocities at the virial radius, $v_c$, of the order $ \lesssim $ 50 km s$^{-1}$; haloes with $v_c \lesssim 30$ km s$^{-1}$ may be inhibited from star formation by the UV background radiation (e.g. \citealt{bib:Rees86}; \citealt{bib:Efstathiou92}; \citealt{bib:Thoul96}; \citealt{bib:Gnedin00}).  Thus many subhaloes may remain dark, and the fraction of bright subhaloes will not increase significantly.  Increasing the resolution of the simulation would also mean that some of the `orphan' galaxies would be resolved.  Since only `orphan' galaxies are allowed to merge with the central galaxy, increasing the resolution may prolong the lifetime of some of our `orphan' galaxies.  To fully understand the impact of this effect a more quantitative analysis is required. Clearly, a firm conclusion can only be reached with higher-resolution simulations with realistic treatment of the gas processes.

\section{Summary and Discussion}
\label{summary}
In summary, for the CLASS survey, approximately 5 of the 22 primary lensing galaxies appear to have a faint
companion within the projected central 5 $ h^{-1}$ kpc. The companions have luminosities of about $2 - 40$\% of the primary galaxy. 
We have studied host galaxies covering a comparable range of
luminosities and host-to-satellite separations to the CLASS lenses, and
found that the predicted fraction of galaxy-(group-) sized haloes hosting central luminous satellites ($\sim$ 3\% (6\%)  at $z = 0$; $\sim$ 11\% (17\%) at $z = 1$) is slightly lower than (but possibly consistent with) the observed value. 

While this fraction is largely independent of galaxy type, it is shown to increase with redshift. 
The Poisson probability of detecting luminous substructure in 5 out of 22 lenses, given that $3\%$ of haloes host luminous substructure, is $\sim 5 \times 10^{-4}$.  If $17\%$ of haloes host luminous substructure, the probability of such a detection is $\sim 0.14$. Our prediction of the redshift and mass dependence appears to be roughly consistent with the data: three lenses with luminous satellites are in groups (see Section \ref{sec:observations}), and all
appear have redshifts close to 1, higher than the median redshift ($\sim$ 0.6) of all CLASS lenses (see the right panel of Fig. \ref{obs}).  
One possibility that we have not considered, is whether lensing galaxies are biased tracers of substructure; such bias may arise if substructure enhances the lensing cross-sections significantly. Previous studies, on cluster scales, for giant arcs indicates that the bias is small (\citealt{bib:Hennawi07}); it remains to be seen whether this holds true for galaxy-scale lenses.  Observationally, the Sloan Lens ACS Survey (SLACS) seems to indicate that the lensing galaxies at z $\sim$ 0.2 are typical early-type galaxies (\citealt{bib:Treu08}). Another possibility is that some of the luminous `satellites' are not associated with lensing galaxies at all, but just happen to be along the line of sight (\citealt{2005ApJ...629..673M}).

\cite{bib:Shin08} recently studied the effect of satellite galaxies on gravitational lensing flux ratios using analytic expressions for the
host potential and the satellite galaxies.  They use a spherically
symmetric galaxy distribution, and assume that the three-dimensional
number density falls off like $r^{-3.5}$, comparable to the Milky
Way.  They show that the probability of a finding a large dwarf is about
10\%  within two Einstein radii and about 3\% within one Einstein
radius.  We find that our $z = 0$ results are consistent with this. 

The Millennium Simulation assumes a power-spectrum normalisation of $\sigma_8 = 0.9$, slightly higher than the latest \emph{WMAP} five-year result (\citealt{bib:wmap5}), where $\sigma_8 = 0.8$. A lower value of $\sigma_8$ will mean that haloes are expected to form later and be less concentrated.  However, the impact of this parameter on our results is complicated.  The semi-analytic models allow some fine-tuning of parameters to match observations.  For example, \cite{bib:Wang08} found no significant difference in the galaxy populations (at the redshift range relevant here) created from semi-analytic models based on the \emph{WMAP} one-year
(\citealt{bib:wmap1_03}) and \emph{WMAP} three-year
(\citealt{bib:Spergel07}) $\sigma_8$ values of 0.9 and 0.722, provided suitable galaxy formation parameters were chosen (the difference becomes significant at high redshift).  We find that our results do not change significantly when based on the \emph{WMAP}3 galaxy catalogue produced by \cite{bib:Wang08} when the same merger timescale is adopted (as in their model C).  However, in their model B (which has the same star formation efficiency but a shorter merger timescale than the \citealt{bib:DeLucia07} catalogue) we find a factor of $\sim$ 2 fewer haloes with central substructure.

To summarise, while we find that the fraction of luminous satellites in group-sized haloes at $ z \sim 1$ is roughly consistent with the observational data we caution that a firm conclusion can only be reached with higher-resolution simulations involving a realistic treatment of the gas processes.  At the same time, a larger sample of gravitational lenses will also be beneficial to constrain these models and allow more definitive conclusions on the properties of the substructure to be made.

\section*{Acknowledgements}
We thank Ian Browne, Neal Jackson, Liang Gao, Peter Schneider, Dandan Xu and Simon White for useful discussions and Gerard Lemson and Jie Wang for providing the galaxy catalogues.
The Millennium Simulation databases used in this paper and the web application providing online access to them were constructed as part of the activities of the German Astrophysical Virtual Observatory.
SEB acknowledges the support provided by the EU Framework 6 Marie Curie Early Stage
Training Programme under contract number MEST-CT-2005-19669 ``ESTRELA". 
SM acknowledges travel support from the European Community's Sixth Framework
Marie Curie Research Training Network Programme, contract number MRTN-CT-2004-505183 ``ANGLES''.

\bibliographystyle{mn2e}
\bibliography{ms}

\label{lastpage}

\end{document}